\documentclass[aps,reprint,prl,superscriptaddress,floatfix]{revtex4-2}
\usepackage[utf8]{inputenc}
\usepackage[T1]{fontenc}
\usepackage{lmodern}
\usepackage{amsmath,amssymb}
\usepackage{graphicx}
\usepackage{color}
\usepackage[dvipsnames]{xcolor}
\usepackage{bm}
\usepackage{physics}
\usepackage[colorlinks=true,linkcolor=blue,citecolor=blue,urlcolor=blue]{hyperref}

\newcommand{\kv}{\bm{k}}
\newcommand{\kvh}{\hat{\bm{k}}}

\newcommand{\pvh}{\hat{\bm{p}}}
\newcommand{\Rv}{\bm{R}}

\newcommand{\rv}{\bm{r}}

\renewcommand{\pv}{\bm{p}}

\newcommand{\zerov}{\vb 0}

\newcommand{\edit}[1]{{\leavevmode\color{black}{#1}}}

\begin{document}

\title{Microscopic Model of Exciton Polarons in 2D Wigner Crystals}
\author{Haydn S. Adlong}
\affiliation{Institute for Quantum Electronics, ETH Zürich, Zürich, Switzerland}
\affiliation{Institute for Theoretical Physics, ETH Zürich, Zürich, Switzerland}
\author{Eugen Dizer}
\affiliation{Institut für Theoretische Physik, Universität Heidelberg, 69120 Heidelberg, Germany}
\author{Richard Schmidt}
\affiliation{Institut für Theoretische Physik, Universität Heidelberg, 69120 Heidelberg, Germany}
\author{Ata\c{c} \.{I}mamo\u{g}lu}
\affiliation{Institute for Quantum Electronics, ETH Zürich, Zürich, Switzerland}
\author{Arthur Christianen}
\affiliation{Institute for Quantum Electronics, ETH Zürich, Zürich, Switzerland}
\affiliation{Institute for Theoretical Physics, ETH Zürich, Zürich, Switzerland}
\date{\today}

\begin{abstract}
Monolayer transition-metal dichalcogenides (TMDs) provide a platform for realizing Wigner crystals and enable their detection via exciton spectroscopy. We develop a microscopic theoretical model for excitons interacting with the localized electrons of the Wigner crystal, including their vibrational motion. In addition to the previously observed exciton–Umklapp feature, the theory reproduces and explains the higher-band attractive-polaron resonances recently reported experimentally. Our model further uncovers that the appearance of two equal-strength and parallel attractive polarons, as commonly observed in WSe$_2$ and WS$_2$, is a signature of strong correlations in the electronic system.
Altogether, our results demonstrate that accounting for electronic interactions is essential to reproduce and interpret the exciton–polaron spectra of TMDs.
\end{abstract}

\maketitle

Wigner crystals are a paradigmatic example of strongly correlated electron systems \cite{wigner:1934}, where Coulomb interactions dominate over the kinetic energy and drive the spontaneous formation of an ordered electron lattice. Although predicted almost a century ago, interest in Wigner crystals has recently been revived due to their realization in two-dimensional semiconductors \cite{smolenski:2021,zhou:2021,xiang:2025,sung:2025,zhang:2025,chen:2025}, particularly transition-metal dichalcogenides (TMDs). Here, the strong Coulomb interactions and the low and tunable charge carrier densities give unprecedented experimental access to this correlated phase and its transitions. 

Exciton spectroscopy has proven to be a sensitive probe of charge order in TMDs, where exciton-electron interactions ensure that the formation of a periodic charge distribution gives rise to a periodic potential for excitons. Consequently, excitonic Umklapp resonances \cite{Shimazaki2021} with an energy shift proportional to the electron density emerge in the reflection spectrum \cite{ smolenski:2021,sung:2025,kiper:2025}, signaling the formation of a Wigner crystal. Recently, electron density-dependent, weak resonances have been reported above the attractive polaron lines for WSe$_2$ monolayers when the electrons form Wigner crystals \cite{zhang_2:2025,wang:2025,Liu2026}. Theoretical modeling is essential to understand these exciton spectra and the underlying role of the electronic interactions. 

While a comprehensive theoretical description of polarons in Wigner crystals is still missing, the simpler problem of an excitonic impurity in a non-interacting Fermi sea has been thoroughly explored~\cite{chevy:2006,schmidt:2012, massignan:2025}.
It turns out that exciton spectra in MoSe$_2$ are well described with a simple variational Ansatz, describing the dressing of the exciton with only a single excitation from the Fermi sea \cite{sidler:2017,efimkin:2017,glazov:2020, Huang2023, Mulkerin:2023}. When charge carriers are injected, the neutral exciton resonance shifts to higher energy, forming a repulsive polaron (RP), while an attractive polaron (AP) emerges at lower energy. \edit{For excitons interacting with electrons in lattices, as relevant in Moir{\'e} systems, novel peaks can emerge in the spectrum \cite{Amelio2024,amelio:2024,pichler2025,kiper:2025}}. 

\begin{figure}[b]
    \centering
    \includegraphics[width=0.99\linewidth]{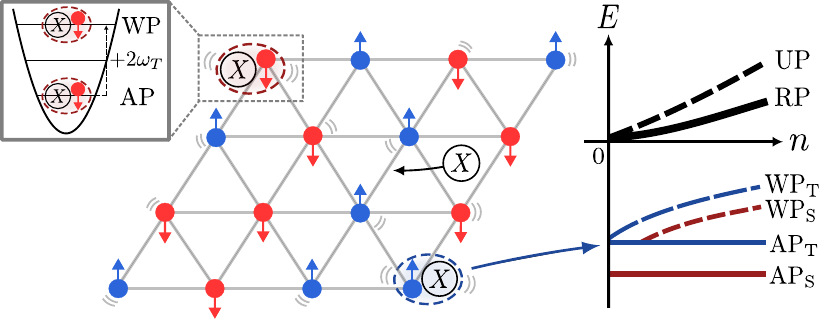}
    \caption{Illustration of excitons ($X$) interacting with a spin-disordered Wigner crystal of electrons in WSe$_2$ and the resulting exciton reflection spectrum as a function of electron density $n$ \cite{zhang_2:2025,wang:2025}. The electrons are modeled as distinguishable particles subject to effective harmonic potentials originating from the restoring Coulomb forces of the Wigner crystal lattice. The exciton can form singlet (S) or triplet (T) trions through interactions with spin-down or spin-up electrons, giving rise to attractive polarons AP$_\text{S}$ and AP$_\text{T}$. Above the attractive polarons, Wigner polarons (WPs) emerge: APs dressed with vibrational excitations of the Wigner crystal with frequency $\omega_T$.}
    \label{fig:intro}
\end{figure}

Here we develop a microscopic theory for excitons interacting with a Wigner crystal, which captures both the RP and AP, as well as the additional resonances that appear when the electrons crystallize (see Fig.~\ref{fig:intro}).
Our model explains the recently observed novel peak emerging above the attractive polaron \cite{zhang_2:2025,wang:2025,Liu2026} as originating from a vibrational
excitation of the Wigner crystal on top of the attractive
polaron. We therefore refer to this feature as a Wigner
polaron.  By contrast, previous models of the Umklapp feature in Wigner crystals have relied on a completely static picture of the electrons \cite{smolenski:2021, vantuan:2023}, and excluded both the attractive and the Wigner polaron.

Remarkably, our model also resolves a long-standing discrepancy between theory and experiment in the spectra of electron-doped WSe$_2$ \cite{courtade:2017,vantuan:2022} and WS$_2$ \cite{zipfel:2020}: experimentally, the singlet and triplet trions evolve into attractive polarons of comparable strength. By contrast, previous polaron theories predict strong hybridization between these peaks, resulting in the transfer of all the oscillator strength to the lower energy peak. Here, we show that the electronic repulsion naturally suppresses this hybridization, demonstrating that the experimental observations can be explained by electronic correlations.

In the following, we introduce the theoretical framework that describes excitons interacting with Wigner crystals in arbitrary spin configurations. We then present the key results: the emergence of higher-band attractive polarons and the absence of hybridization between the two attractive polaron branches in WSe$_2$.

\paragraph*{Model.---}

We consider the interaction of a point-like exciton with a charge-doped TMD monolayer deep in the Wigner-crystal regime (see Fig.~\ref{fig:intro}). Concretely, we focus on an electron-doped WSe$_2$ monolayer encapsulated in hexagonal boron nitride, although our method applies straightforwardly to hole doping and other TMDs. In the deep Wigner crystal limit at low electron density, the electrons are spatially well separated. The exchange interactions between the electrons are therefore negligible, allowing us to treat them as distinguishable particles. We take into account the Coulomb interaction on the mean-field level, where every electron is subject to the Hartree potential of all others. We employ a harmonic approximation for this potential, with the electronic harmonic frequency given by $\omega_e$.
The Hamiltonian in first quantization ($\hbar = 1$) with momentum and position operators $\hat{\bm{P}}_X$ and $\hat{\bm{R}}_X$ for the exciton, and $\hat{\bm{P}}_{i,e}$ and $\hat{\bm{R}}_{i,e}$ for the electron with equilibrium position $\bm{a}_i$ reads
\begin{align} \label{eq:Hamiltonian}
    \hat{\mathcal{H}}&=\frac{|\hat{\bm{P}}_X|^2}{2 m_X} + \sum_i \left[ \hat{{H}}_{i,e}   + g(\hat{\sigma}_{i,e})  \delta(\hat{\bm{R}}_{i,e}-\hat{\bm{R}}_X)  \right] \,, \\
    \hat{{H}}_{i,e}&=\frac{|\hat{\bm{P}}_{e,i}|^2}{2 m_e} +\frac{1}{2}m_e \omega_e^2 (\bm{\hat{R}}_{i,e}-\bm{a}_i)^2 \,. \label{eq:Hamel}
\end{align}
Here $m_X = 0.81 m_0$ and $m_e=0.54m_0$ are the exciton and electron masses, respectively, in terms of the bare electron mass $m_0$ \footnote{We take the masses as chosen in Ref.~\cite{christianen:2025}, primarily based on Ref.~\cite{goryca:2019}}. 

We employ contact interactions between the exciton and the electrons~\cite{Fey:2020,Efimkin:2021}, where the interaction strength $g(\hat{\sigma}_{i,e})$ between the electrons and the exciton depends on the electronic spin/valley, with operator $\hat{\sigma}_{i,e}$. When the exciton is excited in the $K$-valley, it can create a singlet or triplet trion with an electron
from the lower $K$- or $K'$-conduction band, respectively \cite{courtade:2017,christianen:2025}. We label the attractive polarons originating from these trions as the AP$_\mathrm{S}$, AP$_\mathrm{T}$, see Fig.~\ref{fig:intro}. Due to exchange interactions, these trions have a different binding energy (35 versus 29 meV). We thus choose values of $g(\sigma_{i,e}=\pm \frac{1}{2})$ to reproduce the experimentally observed trion energies (for details, see the Supplemental Material (SM)~\cite{supmat}).

The harmonic frequency of the electrons can be found analytically: 
\begin{equation} \label{Eq:omegae}
\omega_e = \sqrt{\frac{e^2 \zeta}{8 \pi \epsilon_0 \kappa a^3 m_e}} =\sqrt{\frac{3\sqrt{3}e^2 \zeta }{64 \pi  \epsilon_0 \kappa m_e}} n^{3/4}.
\end{equation}
This expression follows from the sum of the Coulomb potentials from all sites. Here $\kappa$ is the dielectric constant, $a$ the Wigner crystal lattice constant, and $\zeta$ a numerical factor determined by the form of the electron–electron interaction. For unscreened Coulomb interactions we find $\zeta \simeq 11.034$, which is reduced in the presence of gate screening. Because the distances between the electrons in the Wigner crystals are large, the effects of deviations from the Coulomb interactions at short distances (such as in the Rytova-Keldysh model \cite{rytova:1967,keldysh:1979}) are small.

\paragraph*{Exciton spectral function.---}
The experimental signal in optical reflection spectroscopy is proportional to the zero-momentum exciton spectral function $A(\omega)$, which we therefore aim to compute. The spectral function can be expressed in terms of the exciton Green's function $G$ as
\begin{align}
    A(\omega) &= -\frac{1}{\pi} \mathrm{Im}\left[ G_{\zerov \zerov}(\omega + i \Gamma) \right] \,, \\
    G_{\kv \kv'}^{-1} (\omega) &= \omega - \frac{|\kv|^2}{2m_X} \delta_{\kv \kv'} - \Sigma_{\kv \kv'} (\omega) \,.
\end{align}
We introduce a finite broadening $\Gamma = 1$ meV, which is chosen to roughly match the finite linewidth of the exciton in experiment and also ensures that we choose the retarded Green's function. The non-trivial part in these expressions is the exciton self-energy $\Sigma_{\kv \kv'} (\omega)$, which we will calculate in the following. 

We focus on the zero temperature case, so that all electrons are initially in the ground-state of their harmonic oscillators. We then divide the total Hilbert space into two subspaces: the subspace where all electrons are in their lowest harmonic oscillator states, and the subspace where at least one electron is excited to a higher level by the exciton. Defining the projector $\hat{\mathcal{P}}$, where all electrons are projected into their ground state, the Hamiltonian of Eq.~\eqref{eq:Hamiltonian} can thus be decomposed as 
\begin{align}
\hat{\mathcal{H}} &=\hat{\mathcal{P}}\hat{\mathcal{H}}\hat{\mathcal{P}}+\left[\hat{\mathcal{P}}\hat{\mathcal{H}}(\hat{\mathcal{I}}-\hat{\mathcal{P}})+h.c.\right]+(\hat{\mathcal{I}}-\hat{\mathcal{P}}) \hat{\mathcal{H}} (\hat{\mathcal{I}}-\hat{\mathcal{P}}) \notag \\
&\approx \hat{\mathcal{H}}^{(0)} + \hat{\mathcal{V}} + \hat{\mathcal{V}}^{\dagger} + \hat{\mathcal{H}}^{(1)} \,, \label{eq:Hsubspaces}
\end{align}
To make the calculation tractable, we assume that the exciton can only excite one electron at a time, leading to the approximation in the second line, where $\hat{\mathcal{H}}^{(1)}$ represents the Hamiltonian for a single electronic excitation and $\hat{\mathcal{V}}$ represents the coupling between the zero- and one-excitation subspaces. This is similar in spirit to the Chevy ansatz for the Fermi polaron, which analogously restricts the interactions to a single particle-hole excitation of the Fermi sea~\cite{chevy:2006,schmidt:2012,massignan:2025}.

The $\hat{\mathcal{H}}^{(0)}$ term in Eq.~\eqref{eq:Hsubspaces} is given by
\begin{equation}
    \hat{\mathcal{H}}^{(0)}= \frac{|\hat{\bm{P}}_X|^2}{2 m_X}  + U(\hat{\bm{R}}_X) \,,
\end{equation}
where $U(\hat{\bm{R}}_X)$ represents the Hartree potential the exciton is subject to from the electrons in their lowest harmonic oscillator states. The operators $\hat{\mathcal{V}}$ and $\hat{\mathcal{H}}_{1}$ can decomposed into the contributions from the different sites:
\begin{align}
    \hat{\mathcal{V}} &=\sum_j \sum_{\bm{k},\lambda} e^{-i \bm{k} \bm{a}_j} V_{\bm{k},\lambda}(\sigma_j) |\bm{k}\rangle \langle j; \lambda | \,, \label{Eq:VSite} \\
    \hat{\mathcal{H}}^{(1)} &=\sum_j \sum_{\lambda,\lambda'} H^{(1)}_{\lambda,\lambda'}(\sigma_j) |j; \lambda\rangle \langle j; \lambda' | \,. \label{Eq:H1Site}
\end{align}
Here $|\bm{k}\rangle$ correspond to excitonic momentum states, and $|j; \lambda\rangle$ is the basis representing correlated states of the exciton and an excited electron on site $j$~\cite{supmat}. Importantly $H^{(1)}$ is diagonal in the electronic site and $V$ and $H^{(1)}$ depend on the site only via the spin index $\sigma_j$. 

As a result, the self energy of the exciton can be written as follows
\begin{equation} \label{eq:Eself}
        \Sigma_{\bm{k}\bm{k}'}(\omega) =  U_{\bm{k}\bm{k}'} + \sum_j \bm{V}_{\bm{k}}(\sigma_j) \frac{e^{i (\bm{k}'-\bm{k}) \bm{a}_j}}{\omega-\bm{H}^{(1)}(\sigma_j)} \bm{V}^{\dagger}_{\bm{k}'}(\sigma_j) \,.
\end{equation}
In this form, the self-energy and the spectral function can simply be computed by summing over sites. In this formalism, the two-body electron-exciton problem is solved independently on every site, rendering the polaron calculation remarkably simple and efficient for any spin configuration.

\begin{figure}[!t]
    \centering
    \includegraphics[width=\linewidth]{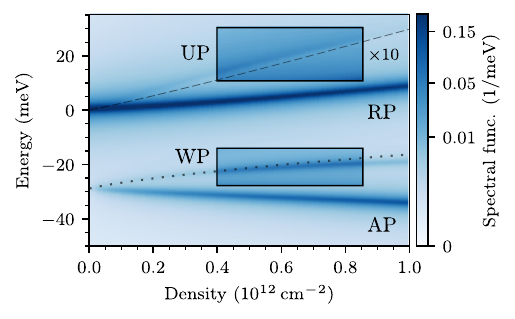}
    \caption{Polaron spectrum as a function of electron density $n$, corresponding to an exciton interacting with a spin-polarized Wigner crystal of electrons in WSe$_2$. Aside from the attractive polaron (AP) and repulsive polaron (RP) peaks, the Umklapp resonance (UP) and the Wigner polaron (WP) are also observed. The black dotted line indicates the energy $E_{\text{AP}}+2\omega_T$ (see main text) and the black dashed line indicates the expected Umklapp energy $E_{\mathrm{UP}}= E_{\mathrm{RP}} +\frac{4 \pi^2 n}{\sqrt{3}m_X}$.} %
    \label{fig:polaron_spinpol}
\end{figure}
\paragraph*{Results: spin-polarized case.---} We first consider the case where the bath is fully spin-polarized, as one would find in the presence of a strong magnetic field.
We show the spectral function $A(\omega)$ as a function of the electron density $n$ in Fig.~\ref{fig:polaron_spinpol}.

\begin{figure*}[!t]
    \centering
    \includegraphics[width=\linewidth]{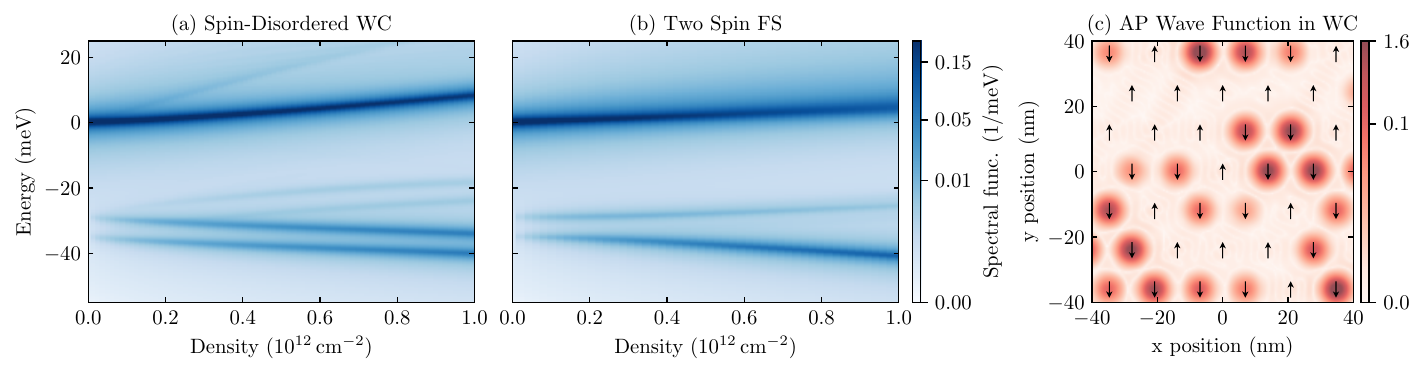}
    \caption{a-b) Polaron spectrum in WSe$_2$ as a function of electron density $n$, corresponding to an exciton interacting with a) a spin-disordered Wigner crystal, b) a mixture of two non-interacting Fermi seas. c) The singlet (ground state) attractive polaron (AP) wave function as a colormap in the Wigner crystal (WC) case at $n= 6 \times 10^{11}$cm$^{-2}$, showing it is localized only on the spin-down sites. The plotted wave function is only the excitonic portion of the full AP wave function in the zero excitation subspace (see $\alpha$ component of the wave function discussed in the Eq.~(S2) of the SM~\cite{supmat}).}
    \label{fig:polaron_spindisorder}
\end{figure*}

The polaron spectrum shows the usual attractive polaron (AP) and repulsive polaron (RP) branches, as well as two branches that appear specifically when the electrons form a Wigner crystal. The Umklapp polaron (UP) branch lies above the repulsive polaron as discussed and experimentally observed previously \cite{smolenski:2021}. However, our model also captures the Wigner polaron (WP) branch, as observed in recent experiments \cite{zhang_2:2025,wang:2025}.

The Wigner polaron can be interpreted as originating from a higher harmonic oscillator state of the trion. If we neglect the coupling between the center-of-mass and the internal motion, the trion vibrational frequency is simply given by $\omega_T=\sqrt{\frac{m_e}{m_T}} \omega_e$, where $m_T=m_X+m_e$ is the trion mass. The Wigner polaron branch appears at $2 \omega_T$ higher energy than the attractive polaron branch, as indicated by the black dotted line in Fig.~\ref{fig:polaron_spinpol}. The splitting between the AP and WP therefore approximately follows an $n^{3/4}$ scaling \cite{zhang_2:2025}, in contrast to the linear $n$ scaling of the splitting between the Umklapp peak and the repulsive polaron.

The Wigner-polaron oscillator strength is small --- around 10\% of that of the attractive polaron. Its spectral weight follows from projecting the electron’s lowest harmonic-oscillator state onto the vibrational eigenstates of the trion.
Due to the symmetry of the harmonic oscillator wave functions, only the harmonic oscillator states of the trion with zero angular momentum have overlap with the initial electron state. Since the harmonic oscillator length $l_e=1/\sqrt{m_e \omega_e}$ --- setting the extent of the harmonic oscillator wave functions --- scales as $l_e \sim m_e^{1/4}$, it is only 25\% different between the electron and trion. 
Consequently, the wave function overlap is dominated by the lowest trion harmonic oscillator state.

\paragraph*{Results: spin-unpolarized case.---} We now move to the scenario without external magnetic field and we assume that the electron spins are disordered.
We model the spin-disordered regime by considering a supercell with a random spin configuration and periodic boundary conditions. If the supercell is large enough, the polaron spectrum is indistinguishable from the case where the spin configuration is random across the whole lattice.

The polaron spectrum, shown in Fig.~\ref{fig:polaron_spindisorder}a), features two parallel and equal strength AP lines (AP$_\mathrm{S}$ and AP$_\mathrm{T}$), each accompanied by a WP line (WP$_\mathrm{S}$ and WP$_\mathrm{T}$). The AP-WP splitting is the same for the AP$_\mathrm{S}$ and AP$_\mathrm{T}$.
By contrast, there is only one repulsive polaron and one Umklapp peak.
This is expected since the exciton scattering with either of the two types of electrons is similar.
Thus, despite the fact that both the WP and Umklapp peaks arise due to the presence of Wigner crystals, our findings highlight that they have a conceptually different origin.

When comparing the polaron spectrum in a Wigner crystal to the polaron spectrum in a non-interacting Fermi sea of both species, as shown in Fig.~\ref{fig:polaron_spindisorder}b), striking differences become evident~\footnote{The case of two non-interacting Fermi seas is calculated within the Chevy approximation~\cite{chevy:2006}, in which the self-energy contributions from the two seas are additive.}.
In addition to the WP peaks disappearing, the attractive polarons display manifestly different properties.
In the case of a non-interacting Fermi sea the two APs strongly hybridize, leading to a transfer of oscillator strength to the lowest AP and level repulsion between the two states. In the Wigner crystal, this hybridization is suppressed and the two AP features are parallel in energy and have comparable oscillator strength.

To explain the lack of hybridization between the AP$_\mathrm{S}$ and AP$_\mathrm{T}$ states in the Wigner crystal case on the microscopic level, we show the AP$_\mathrm{S}$ wave function in Fig.~\ref{fig:polaron_spindisorder}c).
Here we see that the wave function is localized on only one of the spin species, hindering hybridization. This is only possible when the spins are spatially separated, and when the kinetic energy cost for the exciton to localize on one spin type is smaller than the gap between the two trion energies.
This explanation implies that the precise spin configuration does not play a crucial role and we have verified that even
\edit{in more ordered systems, such as large ferromagnetic domains or a striped phase, the polaron spectra remain the same.}
In the SM, we show that the hybridization between the two APs re-emerges in our model when the harmonic potentials for the excitons are relaxed, by increasing the dielectric constant~\cite{supmat}. This confirms that the spatial overlap between the different spin electronic wave functions determines whether the resonances hybridize or not.

Finally, an important question is how the nature of the Wigner crystal phonons affects the WP features. One key approximation in our model is that we neglect the correlations between electrons, which makes the phonons localized. In the SM, we consider a more realistic phonon spectrum of a finite size Wigner crystal, where one electron is replaced by a trion~\cite{supmat}. We find that a phonon resonance around $\omega_T$ still appears around the site of the trion, but with a significant decay width and energy shift. In the presence of disorder, the homogeneous broadening decreases, since the phonons are more localized, but this is replaced with inhomogeneous broadening of approximately the same magnitude. Fully including these more realistic Wigner crystal phonon properties in the polaron calculation presents an interesting direction for future research.

\paragraph*{Comparison with experiment.---} The recently observed experimental spectra in WSe$_2$ \cite{zhang_2:2025,wang:2025,Liu2026} agree well with our model. Consistent with our model, two APs and WPs were observed and only one exciton and Umklapp peak. Moreover, when the system is optically \cite{zhang_2:2025,wang:2025} or magnetically \cite{wang:2025,Liu2026} spin-polarized, one of the APs as well as one of the WPs disappear. This implies that every WP is directly tied to one AP, as follows naturally from our interpretation. The splitting between the WPs and the APs is also of the same order, yet about a factor 1.5 larger in experiment \cite{zhang_2:2025}. One difference between theory and experiment is that the attractive polarons in the experiment barely shift in energy when the electron density is increased, while our theory predicts a clear redshift. This difference has been observed universally, and has been attributed to a combination of band-gap renormalization and the reduction of exciton binding \cite{sidler:2017,Fey:2020}. Importantly, however, the experimental difference in energy between the AP and RP is generally well reproduced theoretically.

One remaining puzzle is that the WP peak is only experimentally observed in electron doped WSe$_2$, but not for hole doping or in previous MoSe$_2$ devices, even though in both of these cases the presence of Wigner crystals was still evidenced by the excitonic Umklapp peak \cite{smolenski:2021}.
At the same time, the electron Wigner crystal in WSe$_2$ is also remarkably stable, persisting to larger densities and temperatures than in these other settings~\cite{zhang_2:2025,wang:2025,Liu2026}.
Although the microscopic origin of this enhanced stability remains an open question, it is plausible that it also modifies the phonon spectrum, which in turn controls the broadening of the WP peak and thus its visibility.

We point out that the absence of hybridization between attractive polarons --- predicted here as a natural consequence of Wigner crystallization --- has in fact already been observed experimentally in WSe$_2$ \cite{courtade:2017} and WS$_2$ \cite{zipfel:2020}. Remarkably, apart from the newly identified WP peaks, these experimental spectra resemble our model in Fig.~\ref{fig:polaron_spindisorder}a) far more closely than the conventional polaron theory of a homogeneous system in Fig.~\ref{fig:polaron_spindisorder}b), which generically predicts hybridization of the APs.
While this may point to the possibility of Wigner crystals already existing in the previously studied samples, it is also plausible that strong short-range electronic correlations in a Fermi liquid may yield a similar effect even in the absence of a Wigner crystal. This hypothesis is supported by the observation that the parallel AP peaks persist even after WP resonances disappear upon increasing the sample temperature~\cite{zhang_2:2025,wang:2025}, or in samples where the gates are in close proximity of the WSe$_2$ monolayer \cite{dijkstra:2025}, presumably preventing Wigner crystal formation. \edit{To address these and other open questions, it would be interesting to extend our approach to finite temperature and investigate how polaron spectroscopy can be used to characterize the melting of the Wigner crystal.}

\paragraph*{Conclusion.---}
We have theoretically studied exciton-polaron spectra in the presence of Wigner crystals in WSe$_2$.
We model the Wigner crystal as distinguishable electrons in harmonic oscillators and describe the interactions of the exciton with these electrons. Aside from reproducing the usual attractive and repulsive polaron features, we find an excitonic Umklapp peak and a novel Wigner polaron feature, which originates from a higher-harmonic oscillator level of the trion.
Our theoretical method can also be applied to excitons interacting with electrons in moir\'{e} potentials \cite{kiper:2025} and generalized Wigner crystals \cite{xu:2020}.

Interestingly, in a Wigner crystal of both electronic spin species, the two attractive polarons do not hybridize as they would in a mixture of two non-interacting Fermi seas. The lack of hybridization of the two attractive polarons, which is typically observed in WSe$_2$ monolayers, is a major shortcoming of the usual polaron model.
Here we have shown that this effect can be explained by the spatial separation of different spin electrons originating from their repulsive interactions.

\edit{\paragraph*{Note.---}
Two other independent works \cite{wang:2025, Liu2026} have developed alternative understandings of the Wigner polaron. In addition, another theoretical work has recently studied an exciton in an adjacent layer interacting with Wigner crystal phonons~\cite{Nyhegn2025}. We compare the different theoretical models and explanations in the SM~\cite{supmat}.}

\paragraph*{Acknowledgements.---} We thank You Zhou, Tomasz Smole{\'n}ski, and Amine Ben Mhenni for useful discussions. This work was supported by the Swiss National Science Foundation (SNSF) under Grant No. 200021-204076. H.\,S.\,A. acknowledges support from the Swiss Government Excellence Scholarship. E.D. and R.S. acknowledge support from the DFG (German Research Foundation) – ProjectID 273811115 – SFB 1225 ISOQUANT, and Germany’s Excellence Strategy EXC 2181/1 - 390900948 (the
Heidelberg STRUCTURES Excellence Cluster). A.C. was funded by an ETH Fellowship.

\bibliography{references}

@PREAMBLE{
 "\providecommand{\noopsort}[1]{}" 
 # "\providecommand{\singleletter}[1]{#1}%" 
}

@article{goryca:2019,
  title={Revealing exciton masses and dielectric properties of monolayer semiconductors with high magnetic fields},
  author={Goryca, M and Li, Jing and Stier, Andreas V and Taniguchi, Takashi and Watanabe, Kenji and Courtade, Emmanuel and Shree, Shivangi and Robert, Cedric and Urbaszek, Bernhard and Marie, Xavier and Crooker, S. A.},
  journal={Nat. Comm.},
  volume={10},
  number={1},
  pages={4172},
  year={2019},
  publisher={Nature Publishing Group UK London},
doi={10.1038/s41467-019-12180-y}
}

@article{Shimazaki2021,
  title = {Optical Signatures of Periodic Charge Distribution in a Mott-like Correlated Insulator State},
  author = {Shimazaki, Yuya and Kuhlenkamp, Clemens and Schwartz, Ido and Smole\ifmmode \acute{n}\else \'{n}\fi{}ski, Tomasz and Watanabe, Kenji and Taniguchi, Takashi and Kroner, Martin and Schmidt, Richard and Knap, Michael and Imamoglu, Atac},
  journal = {Phys. Rev. X},
  volume = {11},
  issue = {2},
  pages = {021027},
  numpages = {10},
  year = {2021},
  month = {May},
  publisher = {American Physical Society},
  doi = {10.1103/PhysRevX.11.021027},
  url = {https://link.aps.org/doi/10.1103/PhysRevX.11.021027}
}

@article{courtade:2017,
  title={Charged excitons in monolayer {WS}e 2: Experiment and theory},
  author={Courtade, Emmanuel and Semina, M and Manca, Marco and Glazov, MM and Robert, C{\'e}dric and Cadiz, Fabian and Wang, Gang and Taniguchi, T and Watanabe, K and Pierre, M and others},
  journal={Phys. Rev. B},
  volume={96},
  number={8},
  pages={085302},
  year={2017},
  publisher={APS},
doi={10.1103/PhysRevB.96.085302}
}

@article{kiper:2025,
  title={Confined trions and Mott-Wigner states in a purely electrostatic moir{\'e} potential},
  author={Kiper, Natasha and Adlong, Haydn S and Christianen, Arthur and Kroner, Martin and Watanabe, Kenji and Taniguchi, Takashi and {\.I}mamo{\u{g}}lu, Atac},
  journal={Phys. Rev. X},
  volume={15},
  number={1},
  pages={011049},
  year={2025},
  publisher={APS},
doi={10.1103/PhysRevX.15.011049}
}

@article{vantuan:2022,
  title={Six-body and eight-body exciton states in monolayer WSe 2},
  author={Van Tuan, Dinh and Shi, Su-Fei and Xu, Xiaodong and Crooker, Scott A and Dery, Hanan},
  journal={Phys. Rev. Lett.},
  volume={129},
  number={7},
  pages={076801},
  year={2022},
  publisher={APS},
doi={10.1103/PhysRevLett.129.076801}
}

@article{dijkstra:2025,
  title={Ten-valley excitonic complexes in charge-tunable monolayer WSe2},
  author={Dijkstra, Alain and Ben Mhenni, Amine and Van Tuan, Dinh and {\c{C}}etiner, Elif and Schur-Wilkens, Muriel and Kim, Junghwan and Steiner, Laurin and Watanabe, Kenji and Taniguchi, Takashi and Barbone, Matteo and others},
  journal={Nat. Comm.},
  volume={16},
  number={1},
  pages={9743},
  year={2025},
  publisher={Nature Publishing Group UK London},
doi={10.1038/s41467-025-65731-x}
}

@article{smolenski:2021,
  title={Signatures of {W}igner crystal of electrons in a monolayer semiconductor},
  author={Smole{\'n}ski, Tomasz and Dolgirev, Pavel E and Kuhlenkamp, Clemens and Popert, Alexander and Shimazaki, Yuya and Back, Patrick and Lu, Xiaobo and Kroner, Martin and Watanabe, Kenji and Taniguchi, Takashi and others},
  journal={Nature},
  volume={595},
  number={7865},
  pages={53--57},
  year={2021},
  publisher={Nature Publishing Group UK London},
doi={10.1038/s41586-021-03590-4}
}

@article{massignan:2025,
  title={Polarons in atomic gases and two-dimensional semiconductors},
  author={Massignan, Pietro and Schmidt, Richard and Astrakharchik, Grigori E and {\.I}mamoglu, Ata{\c{c}} and Zwierlein, Martin and Arlt, Jan J and Bruun, Georg M},
  journal={arXiv:2501.09618},
  year={2025},
doi={10.48550/arXiv.2501.09618}
}

@article{sidler:2017,
  title={Fermi polaron-polaritons in charge-tunable atomically thin semiconductors},
  author={Sidler, Meinrad and Back, Patrick and Cotlet, Ovidiu and Srivastava, Ajit and Fink, Thomas and Kroner, Martin and Demler, Eugene and Imamoglu, Atac},
  journal={Nat. Phys.},
  volume={13},
  number={3},
  pages={255--261},
  year={2017},
  publisher={Nature Publishing Group UK London},
doi={10.1038/nphys3949}
}

@article{zipfel:2020,
  title={Light--matter coupling and non-equilibrium dynamics of exchange-split trions in monolayer WS2},
  author={Zipfel, Jonas and Wagner, Koloman and Ziegler, Jonas D and Taniguchi, Takashi and Watanabe, Kenji and Semina, Marina A and Chernikov, Alexey},
  journal={J. Chem. Phys.},
  volume={153},
  number={3},
  year={2020},
  publisher={AIP Publishing},
doi={10.1063/5.0012721}
}

@article{efimkin:2017,
  title = {Many-body theory of trion absorption features in two-dimensional semiconductors},
  author = {Efimkin, Dmitry K. and MacDonald, Allan H.},
  journal = {Phys. Rev. B},
  volume = {95},
  issue = {3},
  pages = {035417},
  numpages = {10},
  year = {2017},
  month = {Jan},
  publisher = {American Physical Society},
  doi = {10.1103/PhysRevB.95.035417},
  url = {https://link.aps.org/doi/10.1103/PhysRevB.95.035417}
}

@article{glazov:2020,
  title={Optical properties of charged excitons in two-dimensional semiconductors},
  author={Glazov, Mikhail M},
  journal={J. Chem. Phys.},
  volume={153},
  number={3},
  year={2020},
  publisher={AIP Publishing},
  doi={10.1063/5.0012475}
}

@article{christianen:2025,
title={Asymmetric trions in monolayer transition metal dichalcogenides},
author={Christianen, Arthur and Imamoglu, Atac},
journal={arXiv:2507.16643},
year={2025}
}

@article{chevy:2006,
  title = {Universal phase diagram of a strongly interacting Fermi gas with unbalanced spin populations},
  author = {Chevy, F.},
  journal = {Phys. Rev. A},
  volume = {74},
  issue = {6},
  pages = {063628},
  numpages = {4},
  year = {2006},
  month = {Dec},
  publisher = {American Physical Society},
  doi = {10.1103/PhysRevA.74.063628},
  url = {https://link.aps.org/doi/10.1103/PhysRevA.74.063628}
}

@article{bonsall:1977,
  title = {Some static and dynamical properties of a two-dimensional Wigner crystal},
  author = {Bonsall, Lynn and Maradudin, A. A.},
  journal = {Phys. Rev. B},
  volume = {15},
  issue = {4},
  pages = {1959--1973},
  numpages = {0},
  year = {1977},
  month = {Feb},
  publisher = {American Physical Society},
  doi = {10.1103/PhysRevB.15.1959},
  url = {https://link.aps.org/doi/10.1103/PhysRevB.15.1959}
}

@article{schmidt:2012,
  title = {Fermi polarons in two dimensions},
  author = {Schmidt, Richard and Enss, Tilman and Pietil\"a, Ville and Demler, Eugene},
  journal = {Phys. Rev. A},
  volume = {85},
  issue = {2},
  pages = {021602},
  numpages = {5},
  year = {2012},
  month = {Feb},
  publisher = {American Physical Society},
  doi = {10.1103/PhysRevA.85.021602},
  url = {https://link.aps.org/doi/10.1103/PhysRevA.85.021602}
}

@article{zhang_2:2025,
  title={Wigner polarons reveal Wigner crystal dynamics in a monolayer semiconductor},
  author={Zhang, L. and Gu, L. and Adlong, H. S. and Christianen, A. and Dizer, E. and Ni, R. and Ma, R. and Park, S. and Jang, H. and Taniguchi, T. and Watanabe, K. and Esterlis, I. and Schmidt, R. and Imamoglu, A. and Zhou, Y.},
  journal={arXiv:2512.16631},
  year={2025},
  doi={10.48550/arXiv.2512.16631}
}

@article{wang:2025,
    author = "Wang, L. and Menzel, F. and Pichler, F. and Knüppel, P. and Watanabe, K. and Taniguchi, T. and Knap, M. and Smoleński, T.",
    title = "{Spectroscopy of Wigner crystal polarons in an atomically thin semiconductor}",
    journal = "arXiv:2512.16552",
    eprint = "2512.16552",
    archivePrefix = "arXiv",
    primaryClass = "cond-mat.mes-hall",
    month = "12",
    year = "2025",
    doi = "10.48550/arXiv.2512.16552"
}

@article{zhou:2021,
  title={Bilayer Wigner crystals in a transition metal dichalcogenide heterostructure},
  author={Zhou, You and Sung, Jiho and Brutschea, Elise and Esterlis, Ilya and Wang, Yao and Scuri, Giovanni and Gelly, Ryan J and Heo, Hoseok and Taniguchi, Takashi and Watanabe, Kenji and others},
  journal={Nature},
  volume={595},
  number={7865},
  pages={48--52},
  year={2021},
  publisher={Nature Publishing Group UK London},
doi={10.1038/s41586-021-03560-w}
}

@article{chen:2025,
  title={Terahertz electrodynamics in a zero-field Wigner crystal},
  author={Chen, Su-Di and Qi, Ruishi and Kim, Ha-Leem and Feng, Qixin and Xia, Ruichen and Abeysinghe, Dishan and Xie, Jingxu and Taniguchi, Takashi and Watanabe, Kenji and Lee, Dung-Hai and others},
  journal={arXiv:2509.10624},
  year={2025},
doi={10.48550/arXiv.2509.10624}
}

@article{xiang:2025,
  title={Imaging quantum melting in a disordered 2D Wigner solid},
  author={Xiang, Ziyu and Li, Hongyuan and Xiao, Jianghan and Naik, Mit H and Ge, Zhehao and He, Zehao and Chen, Sudi and Nie, Jiahui and Li, Shiyu and Jiang, Yifan and others},
  journal={Science},
  volume={388},
  number={6748},
  pages={736--740},
  year={2025},
  publisher={American Association for the Advancement of Science},
doi={10.1126/science.ado7136}
}

@article{sung:2025,
  title={An electronic microemulsion phase emerging from a quantum crystal-to-liquid transition},
  author={Sung, Jiho and Wang, Jue and Esterlis, Ilya and Volkov, Pavel A and Scuri, Giovanni and Zhou, You and Brutschea, Elise and Taniguchi, Takashi and Watanabe, Kenji and Yang, Yubo and others},
  journal={Nat. Phys.},
  volume={21},
  number={3},
  pages={437--443},
  year={2025},
  publisher={Nature Publishing Group UK London},
doi={10.1038/s41567-024-02759-8}
}

@article{zhang:2025,
  title={Transport Evidence for Wigner Crystals in Monolayer MoTe2},
  author={Zhang, Mingjie and Wang, Zhenyu and Jiang, Yifan and Liu, Yaotian and Watanabe, Kenji and Taniguchi, Takashi and Liu, Song and Lei, Shiming and Li, Yongqing and Xu, Yang},
  journal={arXiv:2506.20392},
  year={2025},
doi={10.48550/arXiv.2506.20392}
}

@article{wigner:1934,
  title={On the interaction of electrons in metals},
  author={Wigner, Eugene},
  journal={Physical Review},
  volume={46},
  number={11},
  pages={1002},
  year={1934},
  publisher={APS},
doi={10.1103/PhysRev.46.1002}
}

@article{vantuan:2023,
  title = {Excitons in periodic potentials},
  author = {Van Tuan, Dinh and Dery, Hanan},
  journal = {Phys. Rev. B},
  volume = {108},
  issue = {8},
  pages = {L081301},
  numpages = {6},
  year = {2023},
  month = {Aug},
  publisher = {American Physical Society},
  doi = {10.1103/PhysRevB.108.L081301},
  url = {https://link.aps.org/doi/10.1103/PhysRevB.108.L081301}
}

@article{Fey:2020,
  title = {Theory of exciton-electron scattering in atomically thin semiconductors},
  author = {Fey, Christian and Schmelcher, Peter and Imamoglu, Atac and Schmidt, Richard},
  journal = {Phys. Rev. B},
  volume = {101},
  issue = {19},
  pages = {195417},
  numpages = {13},
  year = {2020},
  month = {May},
  publisher = {American Physical Society},
  doi = {10.1103/PhysRevB.101.195417},
  url = {https://link.aps.org/doi/10.1103/PhysRevB.101.195417}
}

@article{Efimkin:2021,
  title = {Electron-exciton interactions in the exciton-polaron problem},
  author = {Efimkin, Dmitry K. and Laird, Emma K. and Levinsen, Jesper and Parish, Meera M. and MacDonald, Allan H.},
  journal = {Phys. Rev. B},
  volume = {103},
  issue = {7},
  pages = {075417},
  numpages = {15},
  year = {2021},
  month = {Feb},
  publisher = {American Physical Society},
  doi = {10.1103/PhysRevB.103.075417},
  url = {https://link.aps.org/doi/10.1103/PhysRevB.103.075417}
}

@article{xu:2020,
  title={Correlated insulating states at fractional fillings of moir{\'e} superlattices},
  author={Xu, Yang and Liu, Song and Rhodes, Daniel A and Watanabe, Kenji and Taniguchi, Takashi and Hone, James and Elser, Veit and Mak, Kin Fai and Shan, Jie},
  journal={Nature},
  volume={587},
  number={7833},
  pages={214--218},
  year={2020},
  publisher={Nature Publishing Group UK London},
doi={10.1038/s41586-020-2868-6}
}

@article{Levinsen:2015,
	Author = {Levinsen, J and Parish, M M},
	Journal = {Annu. Rev. Cold Atoms Mol.},
	Month = {may},
	Pages = {1--75},
	Publisher = {World Scientific},
	Title = {{Strongly interacting two-dimensional Fermi gases}},
	Url = {https://doi.org/10.1142/9789814667746_0001},
	Volume = {3},
	Year = {2015}}

@article{Huang2023,
  title = {Quantum Dynamics of Attractive and Repulsive Polarons in a Doped ${\mathrm{MoSe}}_{2}$ Monolayer},
  author = {Huang, Di and Sampson, Kevin and Ni, Yue and Liu, Zhida and Liang, Danfu and Watanabe, Kenji and Taniguchi, Takashi and Li, Hebin and Martin, Eric and Levinsen, Jesper and Parish, Meera M. and Tutuc, Emanuel and Efimkin, Dmitry K. and Li, Xiaoqin},
  journal = {Phys. Rev. X},
  volume = {13},
  issue = {1},
  pages = {011029},
  numpages = {8},
  year = {2023},
  month = {Mar},
  publisher = {American Physical Society},
  doi = {10.1103/PhysRevX.13.011029},
  url = {https://link.aps.org/doi/10.1103/PhysRevX.13.011029}
}

@article{rytova:1967,
author={Rytova, N.S.},
journal={Proc. MSU Phys. Astron.},
volume={3},
pages={30},
year={1967}
}

@article{keldysh:1979,
author={Keldysh, L.V.},
journal={Pis’ma Zh. Eksp. Teor. Fiz.},
volume={29},
pages={716},
year={1979}
}

@article{Mulkerin:2023,
  title = {Exact Quantum Virial Expansion for the Optical Response of Doped Two-Dimensional Semiconductors},
  author = {Mulkerin, Brendan C. and Tiene, Antonio and Marchetti, Francesca Maria and Parish, Meera M. and Levinsen, Jesper},
  journal = {Phys. Rev. Lett.},
  volume = {131},
  issue = {10},
  pages = {106901},
  numpages = {8},
  year = {2023},
  month = {Sep},
  publisher = {American Physical Society},
  doi = {10.1103/PhysRevLett.131.106901},
  url = {https://link.aps.org/doi/10.1103/PhysRevLett.131.106901}
}

@misc{pichler2025,
    author={Fabian Pichler and Mohammad Hafezi and Michael Knap},
    eprint={2503.07712},
    archivePrefix={arXiv},
    year={2025},
    primaryClass={cond-mat.str-el},
    url={https://arxiv.org/abs/2503.07712}, 
}

@article{Nyhegn2025,
    title = "{An exciton interacting with the phonons of an electronic Wigner crystal}",
    author = {J. H. Nyhegn and E. R. Christensen and G. M. Bruun},
    journal = "arXiv:2512.16888",
    eprint = "2512.16888",
    archivePrefix = "arXiv",
    primaryClass = "cond-mat.str-el",
    month = "12",
    year = "2025",
    doi = "10.48550/arXiv.2512.16888"
}

@article{Liu2026,
    author = "Liu, E. and Wilson, M. and Hu, J. and Zimmerman, A. and Mathew, A. and Ouyang, T. and Shi, A. and Taniguchi, T. and Watanabe, K. and Heinz, T. F. and Chang, Y.-C. and Lui, C. H.",
    title = "{Exciton–polaron Umklapp scattering in Wigner crystals}",
    journal = "arXiv:2601.11914",
    eprint = "2601.11914",
    archivePrefix = "arXiv",
    primaryClass = "cond-mat.mes-hall",
    month = "1",
    year = "2026",
    doi = "10.48550/arXiv.2601.11914"
}

@article{Amelio2024,
  title = {Polaron formation in insulators and the key role of hole scattering processes in band insulators, charge density waves, and Mott transitions},
  author = {Amelio, Ivan and Mazza, Giacomo and Goldman, Nathan},
  journal = {Phys. Rev. B},
  volume = {110},
  issue = {23},
  pages = {235302},
  numpages = {13},
  year = {2024},
  month = {Dec},
  publisher = {American Physical Society},
  doi = {10.1103/PhysRevB.110.235302},
  url = {https://link.aps.org/doi/10.1103/PhysRevB.110.235302}
}

@article{amelio:2024,
  title={Polaron spectroscopy of interacting Fermi systems: Insights from exact diagonalization},
  author={Amelio, Ivan and Goldman, Nathan},
  journal={SciPost Physics},
  volume={16},
  number={2},
  pages={056},
  year={2024}
}

@misc{supmat,
	Note = {See the supplemental material for details of the model and theoretical methods, and further work on the re-emergence of attractive polaron hybridization and the case of a disordered 2D Wigner crystal. This includes Refs.~\cite{Levinsen:2015,bonsall:1977}.}}

\renewcommand{\theequation}{S\arabic{equation}}
\renewcommand{\thefigure}{S\arabic{figure}}
\renewcommand{\thetable}{S\arabic{table}}

\onecolumngrid


\setcounter{equation}{0}
\setcounter{figure}{0}
\setcounter{table}{0}

\clearpage

\section*{SUPPLEMENTAL MATERIAL:\\ ``Theory of exciton polarons in 2D Wigner crystals''}
\setcounter{page}{1}
\begin{center}
Haydn S. Adlong,$^{1,2}$
Eugen Dizer,$^{3}$
Richard Schmidt,$^3$
Atac \.{I}mamo\u{g}lu,$^1$ and
Arthur Christianen,$^{1,2}$ \\
\emph{\small $^1$Institute for Quantum Electronics, ETH Zürich, Zürich, Switzerland}\\
\emph{\small $^2$Institute for Theoretical Physics, ETH Zürich, Zürich, Switzerland}\\
\emph{\small $^3$Institut für Theoretische Physik, Universität Heidelberg, 69120 Heidelberg, Germany}
\end{center}

\section{Model and theoretical methods}

In this section we provide details on our approach to solving the exciton polaron problem with the Hamiltonian:
\begin{equation} 
    \hat{\mathcal{H}}=\frac{\hat{P}_X^2}{2 m_X} + \sum_i \left[ \frac{\hat{P}_{e,i}^2}{2 m_e} + \frac{1}{2}m_e \omega_e^2 (\hat{\bm{R}}_{i,e}-\bm{a}_i)^2+ g(\hat{\sigma}_{i,e})  \delta(\hat{\bm{R}}_{i,e}-\hat{\bm{R}}_X)  \right] \,,
\end{equation}
where the various symbols are discussed in the main text. Before we begin, we point out that all of the following analysis should be understood as being equivalent to using the variational ansatz:
\begin{align} \label{eq:varwf}
    \Psi(\bm{R}_X,\{\bm{R}_{i,e},\sigma_{i,e}\})=\, &\sum_{\bm{k}} \alpha_{\kv} e^{i \kv  \bm{R}_X} \prod_i \phi_0(\bm{R}_{i,e}-\bm{a_i},\sigma_i) \notag \\
    &+  \sum_j \eta_j(\bm{R}_X-\bm{R}_{j,e}, \bm{R}_{j,e}-\bm{a}_j,\sigma_j)\prod_{i\neq j} \phi_0(\bm{R}_{i,e}-\bm{a}_i,\sigma_i) \,,
\end{align}
where $\phi_0$ is the lowest harmonic oscillator eigenfunction, and $\alpha_{\bm{k}}$ and $\eta_j(\bm{R}_X-\bm{R}_{j,e}, \bm{R}_{j,e}-\bm{a}_j,\sigma_j)$ are the variational parameters. In this case $\eta_j(\bm{R_X}-\bm{R}_{j,e}, \bm{R}_{j,e}-\bm{a}_j,\sigma_j)$ describes the correlated state of an electron and an exciton around site $j$. Note, care must be taken that $\eta_j(\bm{R}_X-\bm{R}_{j,e}, \bm{R}_{j,e}-\bm{a}_j,\sigma_j)$ is orthogonal to $\phi_0(\bm{R}_{i,e}-\bm{a}_i,\sigma_i)$, so that the different parts of the variational wave function do not overlap.

While the variational wave function provides a conceptually simple picture, we do not solve the corresponding variational equations directly. In a spin-disordered Wigner crystal, keeping track of all $\eta_j$ components would cause the full wave function to grow exponentially with the number of sites. Instead, we use the same variational subspace to derive the exciton self-energy, which allows us to efficiently incorporate the contribution of each electron without explicitly constructing the full many-body wave function.

We begin our analysis by first rewriting our Hamiltonian in the variational subspace spanned by Eq.~\eqref{eq:varwf}. We will then solve the problem of an exciton and electron interacting on a single site at the origin. This is followed by analyzing the case where the electron is forbidden to occupy the ground state of the harmonic oscillator. This will be seen to be necessary for the same reason that the $\eta_j$ functions must be made orthogonal to $\phi_0$ above. Using the solution to this single-site problem, we then return to calculate the exciton self energy. We conclude this section with a few matrix elements that are used in our procedure.

\subsection{Variational subspace}
We partition our Hilbert space as done in the main text, by introducing the projector that puts all electrons in their ground state:
\begin{align}
    \hat{\mathcal{P}} = \hat{{I}}_X \otimes \left( \bigotimes_i \hat P_i \right) \,,
\end{align}
where $\hat{{I}}_X$ is the identity operator on the exciton space and $\hat P_i$ projects the electron on site $i$ into its ground state. Here, and throughout, calligraphic symbols act on the full many-body space, while regular symbols act on individual particles. This projector allows us to decompose the Hamiltonian according to
\begin{align}
    \hat{\mathcal{H}} &=\hat{\mathcal{P}}\hat{\mathcal{H}}\hat{\mathcal{P}}+\left[\hat{\mathcal{P}}\hat{\mathcal{H}}(\hat{\mathcal{I}}-\hat{\mathcal{P}})+h.c.\right]+(\hat{\mathcal{I}}-\hat{\mathcal{P}}) \hat{\mathcal{H}} (\hat{\mathcal{I}}-\hat{\mathcal{P}}) \,,
\end{align}
where $\hat{\mathcal{I}}$ is the identity on the full many-body space.

Here we make the key approximation of our approach, which is to limit our variational subspace to one where only one electron can be excited out of the ground state harmonic oscillator at a time. \textit{Within this space}, we have
\begin{align} \label{Eq:ProjectorLimited}
    \hat{\mathcal{I}}-\hat{\mathcal{P}} = \mathcal{\hat S}^{(1)} = \sum_j \hat{{I}}_X \otimes \left( \bigotimes_{i\neq j} \hat P_i \right) \otimes \left( \hat{{I}}_j- \hat P_j \right) \,,
\end{align}
where $\hat{{I}}_j$ is the identity operator for the electron on site $j$. Using these projectors the Hamiltonian is then decomposed into four terms,
\begin{align}
    \hat{\mathcal{H}}\approx \hat{\mathcal{H}}^{(0)} + \hat{\mathcal{V}} + \hat{\mathcal{V}}^{\dagger} + \hat{\mathcal{H}}^{(1)} \,,
\end{align}
which we will now define.

The first term is limited to the zero-excitation subspace:
\begin{align}
    \hat{\mathcal{H}}^{(0)} &= \mathcal{P} \hat{\mathcal{H}} \mathcal{P} =  \frac{\hat{P}_X^2}{2 m_X} + N \omega_e + U(\hat R_X) \,,
\end{align}
where $N$ is the number of electrons and $U$ is the Hartree potential that the exciton experiences from all the electrons occupying the ground state. Denoting the ground state of the many-body electronic system by $\ket{\Phi}$ we have
\begin{align}
    U(\hat R_X) = \bra{\Phi} \hat{V}(\sigma_j) \ket{\Phi} \,,
\end{align}
with 
\begin{align}
    \hat{V}(\hat{\sigma}_j) = g(\hat \sigma_j) \delta(\hat{\bm{R}}_{j,e}-\hat{\bm{R}}_X) \,.
\end{align}
The interaction element $\hat{\mathcal{V}}$, which couples the zero excitation to the one excitation subspace is 
\begin{align}
\hat{\mathcal{V} } = \hat{\mathcal{P}}\hat{\mathcal{H}} \mathcal{\hat S}^{(1)} = \sum_j \hat{P}_j \hat{V}(\hat \sigma_j) \left( \hat{\mathcal{I}}_j- \hat P_j \right) \,.
\end{align}
Finally, the term limited to the one-excitation subspace is
\begin{align} \label{Eq:H(1)Sup}
    \hat{\mathcal{H}}^{(1)} = \mathcal{\hat S}^{(1)} \hat{\mathcal{H}} \mathcal{\hat S}^{(1)} = \frac{\hat{P}_X^2}{2 m_X} + (N-1) \omega_e+ \sum_j \left( \hat{\mathcal{I}}_j- \hat P_j \right) \left( \hat{{H}}_{j,e}   + g(\hat{\sigma}_{j,e})  \delta(\hat{\bm{R}}_{j,e}-\hat{\bm{R}}_X) \right) \left( \hat{\mathcal{I}}_j- \hat P_j \right) \,.
\end{align}
Here we ignore the Hartree contribution from the other electrons to the exciton in the one-excitation subspace since the interactions with the other electrons are negligible when the exciton is localized around one site.

To proceed, we require the solution to the exciton-electron scattering problem on every site. We first solve this generically, and then show it can be adapted in the scenario where the electron is forbidden from entering its ground state, as necessitated by the form of Eq.~\eqref{Eq:H(1)Sup}. We will solve both of these problems for an electron at the origin, since the solution at other sites is related by a simple translation.

\subsection{Exciton-electron problem}
We consider the Hamiltonian of an exciton interacting with a harmonically confined electron centered at the origin:
\begin{align}
    \hat H_{\rm Xe} = \frac{|\hat{\bm{P}}_X|^2}{2 m_X} + \frac{|\hat{\bm{P}}_{e}|^2}{2 m_e} +\frac{1}{2}m_e \omega_e^2 \hat{R}_{e}^2+ g  \delta(\hat{\bm{R}}_{e}-\hat{\bm{R}}_X) \,.
\end{align}
Here we write the problem generically with coupling strength $g$ (i.e., we will later reintroduce the spin and site indices). We find that an efficient way to solve this problem is by working in the coordinates of the electron-position ($\bm{r}_e = \Rv_e$) and the relative coordinate $(\rv = \Rv_X - \Rv_e)$. Note that care must be taken when moving to momentum space. In this representation, we denote the conjugate variables to $\rv_e$ and $\rv$ by $\pv$ and $\kv$, respectively. Notice that these momenta relate to the physical exciton and electron momentum by
\begin{align}
    \pv = \bm{P}_X + \bm{P}_e \,, \qquad \kv = \bm{P}_X \,.
\end{align}
In this basis, the Hamiltonian reads
\begin{align}
    \hat{H}_{\rm Xe} = \frac{|\kvh|^2}{2 \mu} + g \delta (\hat{\rv}) + \frac{|\pvh|^2}{2 m_e} + \frac{1}{2} m_e \omega_e^2 |\hat{\rv}_e|^2 - \frac{1}{m_e} \pvh \cdot \kvh \,,
\end{align}
where $1/\mu = 1/m_e + 1/m_X$. In order to diagonalize this Hamiltonian, we first solve the relative and electron problem separately and then use this as a basis to solve the full coupled Hamiltonian. We find that this method allows us to efficiently capture both the bound and scattering states.

The electron problem is simply that of a harmonic oscillator. We expand the solutions in momentum space
\begin{align}
    \ket{n,l} = \int p \, dp \, \phi^{(l)}_n(p) \frac{e^{il\theta_p}}{\sqrt{2 \pi}} \ket{\pv} \,,
\end{align}
where we used polar coordinates $\pv = (p, \theta_p)$, and our convention is $\bra{\pv} \ket{\pv'} = \delta(\pv - \pv')$. Here, the wave function in momentum space is known analytically:
\begin{align}
    \phi^{(l)}_n(p) = l_e\,
\sqrt{\frac{2n!}{(n + |l|)!}}\;
(pl_e)^{|l|}\,
e^{-\tfrac{1}{2}( pl_e)^2}\,
L_{n}^{(|l|)}\!\big(p^2 l_e ^2\big) \,,
\end{align}
where $l_e=1/\sqrt{m_e \omega_e}$ is the harmonic oscillator length. The corresponding energies are
\begin{align}
    E_{nl} = \omega_e(2n + |l| + 1) \,.
\end{align}
For the relative problem, we similarly use angular momentum modes and expand the solutions according to
\begin{align}
    \ket{\nu, m} = \int k \,dk\, d \theta_k \,  \psi^{(m)}_{\nu}(k) \frac{e^{i m \theta_k}}{\sqrt{2 \pi}} \ket{\kv} \,,
\end{align}
where in polar coordinates $\kv = (k, \theta_k)$. Here, the wave function coefficients $\psi^{(m)}_{\nu}(k)$ are solutions to the linear integral equation
\begin{align}
    E \psi^{(m)}_{\nu}(k) = \frac{k^2}{2 \mu} \psi_\nu^{(m)}(k) + \frac{g}{2 \pi} \delta_{m0} \int_0^\Lambda k' \, dk' \, \psi^{(m)}_{\nu}(k') \,.
\end{align}
Here we observe that the contact interactions only affect the $m=0$ channel; in this channel we impose the necessary ultra-violet cutoff $\Lambda$. The cutoff is related to the trion binding energy $E_T$ by~\cite{Levinsen:2015}
\begin{align}
    \frac{1}{g} = -\frac{1}{2 \pi} \int_0^\Lambda k \, dk \frac{1}{E_T + k^2/2\mu} \,.
\end{align}

In the combined basis $\ket{\nu m; n l}$ the matrix elements of the coupling are given by
\begin{align}
    \frac{1}{m_e}\bra{\nu' m'; n' l'} \pvh \cdot \kvh \ket{\nu m; nl} = \frac{1}{2 m_e} \left( \delta_{m',m'+1} \delta_{l',l-1} +  \delta_{m',m'-1} \delta_{l',l+1} \right) \int p^2 \, dp \, \phi^{(l')}_{n'}(p) \phi^{(l)}_{n}(p) \int k^2 \, dk \, \psi^{(m')}_{\nu'}(k) \psi^{(m)}_{\nu}(k) \,,
\end{align}
where we have used that $\pv \cdot \kv = p k \cos(\theta_p - \theta_k)$. The integral over $p$ can be performed analytically:
\begin{subequations}
    \begin{align}
   \int p^2 \, dp \, \phi^{(l-1)}_{n'}(p) \phi^{(l)}_{n}(p)  &= 
 \frac{1}{l_e} \left( \sqrt{n+|l|} \delta_{n,n'} - \sqrt{n+1} \delta_{n',n+1} \right) \,, \\
  \int p^2 \, dp \, \phi^{(l+1)}_{n'}(p) \phi^{(l)}_{n}(p)  &= 
 \frac{1}{l_e} \left( \sqrt{n+|l|+1} \delta_{n,n'} - \sqrt{n} \delta_{n',n-1} \right) \,.
\end{align}
\end{subequations}

Here we observe that quantum number $l+m$ is conserved, and we thus focus in what follows on states of form $\ket{\nu n m} \equiv \ket{\nu m; n -m}$.

\subsection{Projected exciton-electron problem}

We now reconsider an exciton-electron problem in the scenario where the electron is projected out of the ground state harmonic oscillator. This is non-trivial because the relative exciton-electron distance $\rv$ also depends on $\Rv_e$. To solve this problem, we introduce the projector that puts the electron in the ground state:
\begin{align}
    \hat{P} = \hat{{I}}_X \otimes \dyad{\phi_0^{(0)}} \,,
\end{align}
where $\hat{{I}}_X$ is the identity operator on the exciton space. We can now solve the exciton-electron problem in a projected basis, which is written as
\begin{align}
    \ket{\lambda} = \sum \lambda_{\nu n m} \hat S \ket{\nu n m} \,,
\end{align}
where we have the orthogonal complement operator
\begin{align}
    \hat{S} = \hat{{I}}_2 - \hat{P} \,.
\end{align}
Here $ \hat{{I}}_2 $ is the identity on the full two-body space.

In what follows, we collect the elements $\lambda_{\nu n m}$ into the vector $\bm{\lambda}$ and we define $\bm S$ and $\bm{H}_{Xe}$ as the matrix representations of $\hat S$ and $\hat H_{Xe}$, respectively, in the basis $\ket{\nu nm}$. The matrix elements of $\bm S$ are non-trivial and we derive them in Eq.~\eqref{Eq:MatElP}. The exciton-electron problem is now written as a generalized eigenvalue problem
\begin{align}
    E \bm S \bm \lambda = \bm S \bm H_{\rm Xe} \bm S \bm \lambda \,.
\end{align}

The basis we have employed is almost overcomplete and it is thus numerically unstable (equivalently, one can understand this as having eigenvalues of $\bm S$ that are near zero). We can stabilize the algorithm by using a truncation in the eigenbasis of $\bm S$. That is, we write  $\bm S = \bm U \bm S_D \bm U^\dagger$ and then discard modes associated with small eigenvalues (i.e.,  $\bm U$ is no longer square). We have checked that none of our results depend on this cutoff of eigenvalues. This procedure allows us to rearrange into the form
\begin{align} \label{Eq:etaEigenvalueProb}
    E \bm{\eta} = \bm{\tilde H}_{\rm Xe} \bm{\eta} \,,
\end{align}
where 
\begin{align}
    \bm{\eta} = \sqrt{\bm{S_D}} \bm U ^\dagger \bm \lambda \,, \qquad \bm{\tilde H}_{\rm Xe} =  \sqrt{\bm{S_D}} \bm U^\dagger \bm{H}_{\rm Xe} \bm{U} \sqrt{\bm{S_D}} \,.
\end{align}
By solving Eq.~\eqref{Eq:etaEigenvalueProb} we can transform back to the original basis to obtain
\begin{align} \label{eq:SHS}
    \bm S \bm{H}_{\rm Xe} \bm S = \bm{S} \bm \Lambda  \bm{E_D} \bm{\Lambda}^\dagger \bm{S} \,,
\end{align}
where $\bm \Lambda$ is the matrix of $\lambda$ eigenvectors.

\subsection{Exciton self energy}

We now calculate the exciton self-energy within our approximation of a single excitation at a time. By blockwise inversion, it is:
\begin{align}
    G_{\kv \kv'} (\omega) = \bra{\kv,\Phi} \frac{1}{\omega - \hat{\mathcal{H}}} \ket{\kv', \Phi} \approx  \bra{\kv,\Phi} \frac{1}{\omega - \hat{\mathcal{H}}^{(0)} - \hat{\mathcal{V}} \frac{1}{\omega - \hat{\mathcal{H}}^{(1)}  } \hat{\mathcal{V}}^\dagger  }  \ket{\kv',\Phi} \,,
\end{align}
where $\ket{\kv, \Phi}$ denotes a $\kv$ momentum exciton with all electrons in their ground state. Here we can write the exciton Green's function in terms of the self energy
\begin{align}
    G_{\kv \kv'}^{-1} (\omega) &=\omega -  \frac{|\kv|^2}{2m_X} \delta_{\kv \kv'} - \Sigma_{\kv \kv'} (\omega) \,,
\end{align}
which allows us to identify 
\begin{align}
    \Sigma_{\kv \kv'} &=  U_{\kv \kv'} + \bra{\kv ,\Phi} \hat{\mathcal{V}} \frac{1}{\omega - \hat{\mathcal{H}}^{(1)}  } \hat{\mathcal{V}}^\dagger \ket{\kv', \Phi } \,,
\end{align}
where $U_{\kv \kv'} $ is the Fourier transformed Hartree potential, which we derive at the end of this section in Eq.~\eqref{Eq:UHartreeMatEl}. Here the second part of the self energy, originating from the one excitation sector of the subspace, can be rewritten as
\begin{align}
    \Sigma^{(1)}_{\kv \kv'} \equiv \bra{\kv ,\Phi} \hat{\mathcal{V}} \frac{1}{\omega - \hat{\mathcal{H}}^{(1)}  } \hat{\mathcal{V}}^\dagger \ket{\kv', \Phi } = \sum_j \bra{\kv, \Phi} \hat V(\hat \sigma_j)  \left( \hat{\mathcal{I}}_j- \hat P_j \right) \frac{1}{\omega - \hat{{H}}_j^{(1)}} \left( \hat{\mathcal{I}}_j- \hat P_j \right)  \hat V(\hat \sigma_j) \ket{\kv', \Phi} \,,
\end{align}
where 
\begin{align}
     \hat{{H}}_j^{(1)} = \frac{\hat{P}_X^2}{2 m_X} + (N-1) \omega_e+ \left( \hat{{I}}_j- \hat P_j \right) \left( \hat{{H}}_{j,e}   + g(\hat{\sigma}_{j,e})  \delta(\hat{\bm{R}}_{j,e}-\hat{\bm{R}}_X) \right) \left( \hat{{I}}_j- \hat P_j \right),
\end{align}
which is related to our projected exciton-electron problem via a translation and $g\to g(\hat \sigma_{j,e})$. 

Within this formulation, we thus observe that we only require two exciton-electron calculations: one for the exciton interacting with each spin species of the electron at the origin site. We denote these solutions by their eigenvalue decompositions with $\bm{\Lambda}(\sigma = \pm 1)$, $\bm{E_D}(\sigma = \pm 1)$ as in Eq.~\eqref{eq:SHS}. The one excitation contribution to the self energy is then
\begin{align}
     \Sigma_{\kv \kv'}^{(1)} = \sum_j e^{i (\bm{k}'-\bm{k}) \bm{a}_j}  \bm{V}_{\bm{k}}(\sigma_j) \bm{S} \bm{\Lambda}(\sigma_j) \frac{1}{\omega -(N-1)\omega_e - \bm{E_D}(\sigma_j)} \bm{\Lambda}(\sigma_j)^\dagger \bm{S}\bm{V}^{\dagger}_{\bm{k}'}(\sigma_j) \,,
\end{align}
where $ \bm{V}_{\bm{k}}(\sigma_j)$ is the vector formed from the matrix elements of $\bra{\kv, \phi_0^{(0)}} \hat{V}(\sigma_j) \ket{\nu n m}$, which we derive in Eq.~\eqref{Eq:MatV2}. Here the notation $\ket{\kv, \phi_0^{(0)}}$ denotes an exciton at momentum $\kv$ and an electron in the ground state of the harmonic oscillator at the origin.

\subsection{Matrix elements}
We conclude this section with the three matrix elements that were required in this section. It will be seen that a useful integral is
\begin{align}
    \bar \phi_n^{(m)}(k) &\equiv \int pdp \, d \bar \theta_p \, \phi^{(m)}_n(p) \phi^{(0)}_0 \left(\sqrt{p^2+k^2-2 pk \cos(\bar \theta_p) }\right) \frac{e^{-i \bar \theta_p m}}{2 \pi} \nonumber \\
    &= (-1)^n e^{-\frac{1}{4} k^2 l_e^2} \frac{2^{-| m| -2 n}}{n!} \sqrt{\frac{n!}{(|
   m| +n)!}} (k l_e)^{| m| +2 n} \label{Eq:UsefulInt} \,.
\end{align}

The first matrix element we used was the Fourier transform of the Hartree potential $U(R_X)$. Here we must be careful about taking the thermodynamic limit since we have already implicitly taken the system area $A\to \infty$ in our continuous formulation in momentum space, but we still have a finite number of sites $N$. We therefore reintroduce the area factor in the equations by returning to a discrete formulation in momentum space, then again taking a continuum limit. From this we process, we find
\begin{align}
    \bra{\kv, \phi_0^{(0)}} \hat{V}(\sigma_j) \ket{\kv', \phi_0^{(0)}} &=  \frac{g(\sigma_j)}{A} \int d^2 p\, \frac{\phi_0^{(0)}(|\pv -\kv|)}{\sqrt{2 \pi}}\frac{\phi_0^{(0)}(|\pv -\kv'|)}{\sqrt{2 \pi}} \nonumber \\
    &= \frac{g(\sigma_j)}{ A} \bar \phi_0^{(0)}(|\kv' -\kv|) \nonumber \\
    &= \frac{g(\sigma_j)}{A} e^{ -l_e^2|\kv' -\kv|/4} \,.
\end{align}
We can then sum over each site, which yields
\begin{align}
    U_{\kv \kv'} = \frac{1}{A} \sum_j g(\sigma_j) e^{i (\bm{k}'-\bm{k}) \bm{a}_j}  e^{ -l_e^2|\kv' -\kv|^2/4} \,. \label{Eq:UHartreeMatEl}
\end{align}

The next matrix elements that we require are those of $\hat P$. These are given by
\begin{align}
    \bra{\nu' n' m'} \hat{P} \ket{\nu n m} = \int d \kv \, d \pv' d \pv \, \psi^{(m')}_{\nu'}(k) \psi^{(m)}_{\nu}(k) \phi^{(m')}_{n'} (p') \phi^{(m)}_{n}(p) \phi^{(0)}_0 (| \pv' - \kv| ) \phi^{(0)}_0 (| \pv - \kv| ) \frac{e^{i \theta_k (m-m') +i \theta_{p'} m'-i \theta_{p} m}}{(2 \pi)^3} \,.
\end{align}
By performing a change of variables $\bar \theta_{p} = \theta_p - \theta_k $ (and the same with $p\to p')$ we can make use of Eq.~\eqref{Eq:UsefulInt}, to reduce the integral to a single dimension:
\begin{align}
     \bra{\nu' n' m'} \hat{P} \ket{\nu n m} = \int k dk \, \psi^{(m')}_{\nu'}(k) \psi^{(m)}_{\nu}(k) \bar \phi^{(m')}_{n'} (k) \bar \phi^{(m)}_{n}(k) \,, \label{Eq:MatElP}
\end{align}
which we then perform numerically. The matrix elements of $\hat S$ are then related trivially to $\hat P$.

Finally, the matrix elements of the interaction are
\begin{align}
    \bra{\kv,\phi_0^{(0)}} \hat{V}(\sigma_j)  \ket{\nu n m} &= \frac{g(\sigma_j)}{2\pi \sqrt{A}} \int d^2 k' \, d^2 p \,\frac{ \phi_0^{(0)}(|\pv - \kv|)}{\sqrt{2 \pi}} \psi^{(m)}_{\nu}(k')  \phi^{(m)}_n(p) \frac{e^{i m (\theta_{k'}-\theta_p)}}{2 \pi } \nonumber \\
    &= \frac{g(\sigma_j)}{\sqrt{2 \pi A} }\delta_{m0} \, \bar \phi_n^{(0)}(k) \int k' dk' \, \psi^{(m)}_{\nu}(k') \,, \label{Eq:MatV2}
\end{align}
where the integral over $k'$ is performed numerically.

\section{Re-emergence of attractive polaron hybridization}

\begin{figure*}[!b]
    \centering
    \includegraphics[width=0.52\linewidth]{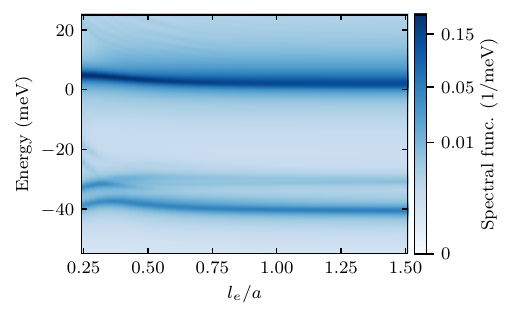}
    \caption{Polaron spectrum in the Wigner crystal model plotted as a function of $l_e/a$ at density $n= 6 \times 10^{11}$cm$^{-2}$. The harmonic oscillator length is changed by varying $\kappa$, smoothly interpolating between the Wigner crystal case (left) and a homogeneous spin mixture of distinguishable electrons (right).} %
    \label{fig:polaron_spindisorder_transition}
\end{figure*}

To study how hybridization progressively emerges as the Wigner crystal breaks down, Fig.~\ref{fig:polaron_spindisorder_transition} shows the effect of suppressing the Coulomb interactions in our model. By enhancing the dielectric constant $\kappa$ we can smoothly interpolate between a Wigner crystal and a gas of distinguishable particles, as characterized by the ratio $l_e/a$. This is of course distinct from the true melting transition, because we do not describe a Fermi liquid state: in our model the electrons remain in harmonic oscillator states for the whole parameter regime. For $\kappa \rightarrow\infty$ ($l_e/a \gg 1)$, we exactly recover the case of an exciton in a non-interacting spin mixture of distinguishable electrons. Our choice of wave function representation makes that we can efficiently describe the polaron spectra across the entire parameter range.

For small $l_e/a$, the WPs first rapidly lose their oscillator strength and redshift, while the APs blueshift. The shift of the APs originates from the increased trion binding for localized electrons \cite{kiper:2025}. Then around $l_e/a \approx 0.35$ the WP$_\mathrm{S}$ merges with the AP$_\mathrm{T}$, leading to an avoided crossing. These states can hybridize because there is a resonant exciton hopping between these states if the WP$_\mathrm{S}$ and the AP$_\mathrm{T}$ have the same energy. However, the hybridization between these states and the lower AP counteracts the reduction in binding from the loss of electron localization, leading to a progressive redshift as $l_e/a$ is further increased. Finally, for $l_e/a >1$ the two lines of the upper AP merge into the single feature as observed in Fig.~\ref{fig:polaron_spindisorder}b) in the main text.

\newpage

\section{Phonon Spectrum and Localization in a Disordered 2D Wigner Crystal}

In this section, we investigate the phonon spectrum of a 2D Wigner crystal with positional disorder. We consider a finite system of $25\times25$ interacting electrons arranged on a triangular lattice with lattice constant $a$. To emulate the electrostatic environment of a larger crystal and prevent edge instabilities, the finite patch is embedded in 22 layers of static background electrons that continue the lattice and provide a stabilizing potential for the mobile electrons. The electron-electron interaction is modeled by a screened Coulomb potential,
\begin{align}
    V(r) = \frac{e^2}{4\pi\epsilon_0} \frac{\exp(-\lambda r)}{\kappa r} \,,
\end{align}
where we choose $\lambda=0.1/a$ as a phenomenological screening length and $\kappa=4.4$, as in the main text. The Yukawa form regularizes the long-range Coulomb interaction while retaining its essential $1/r$ character at short distances, ensuring numerical stability for the finite system and simulating screening effects.

We simulate the disorder by pinning some of the mobile electrons with harmonic trap potentials. We choose a spring constant of $k_{\text{trap}}=2$ meV/nm${}^2$, corresponding to a trap frequency of $\omega_{\text{trap}}=16.8$ meV. The equilibrium positions of the mobile electrons are obtained by relaxing the system through Newton iterations until the total force on each site vanishes. This ensures a mechanically stable configuration around which the phonon modes are computed within the harmonic approximation. The linearized equations of motion for the electron lattice are given in terms of the displacement vector $\bm{U}$ and the force-constant (Hessian) matrix $\Phi$ as~\cite{bonsall:1977}
\begin{align}
    H = \frac{1}{2m_e} \bm{P}^T\bm{P} + \frac{1}{2} \bm{U}^T \Phi \bm{U} \,, \qquad \Phi |n\rangle = m_e \omega^2_n |n\rangle \,.
\end{align}
The phonon frequencies are then obtained from the eigenvalues of the positive-definite matrix $\Phi$ as $\omega_n = \sqrt{E_n/m_e}$, where $m_e=0.54\,m_0$, and $m_0$ is the bare electron mass. Note that $\Phi$ is a $2N\times 2N$ matrix, where $N$ is the number of mobile electron sites, such that the mode number $n\in (0,\dots,2N-1)$. To include a trion site with different mass $m_T=1.35\,m_0$, we modify the force-constant matrix $\Phi$ and consider the mass-weighted eigenvalue problem
\begin{align}
    M^{-1/2} \Phi M^{-1/2} |n\rangle = \omega^2_n |n\rangle \,, \qquad M = \text{diag}(m_e, m_e, \dots, m_T, m_T, \dots, m_e, m_e) \,,
\end{align}
where each site $j$ appears twice for $\nu=x,y$ directions.
Now, the phonon frequencies are simply obtained by $\omega_n = \sqrt{E_n}$. Neglecting the off-diagonal coupling terms, the bare oscillator frequency $\omega_T$ on the trion site is given by
\begin{equation}
    \label{eq:bare_trion_freq}
    \omega_T(n) = \sqrt{\frac{3\sqrt{3}e^2 \tilde{\zeta}}{64 \pi  \epsilon_0 \kappa m_T}}\, n^{3/4} = 1.45 \left(\frac{n}{10^{11}\,\text{cm}^{-2}}\right)^{3/4} \text{meV} \,,
\end{equation}
where we find $\tilde{\zeta}=11.014$ for the above Yukawa potential. This result is modified in presence of disorder and interactions with the other sites, as we will see below.
Figure~\ref{fig:S2_modes_and_configuration} shows an example realization of a relaxed disordered Wigner crystal and its phonon spectrum $\omega_n$ for an electron density of $n=10^{11}$ cm${}^{-2}$.

\begin{figure}[b]
\centering
\includegraphics[width=0.99\textwidth]{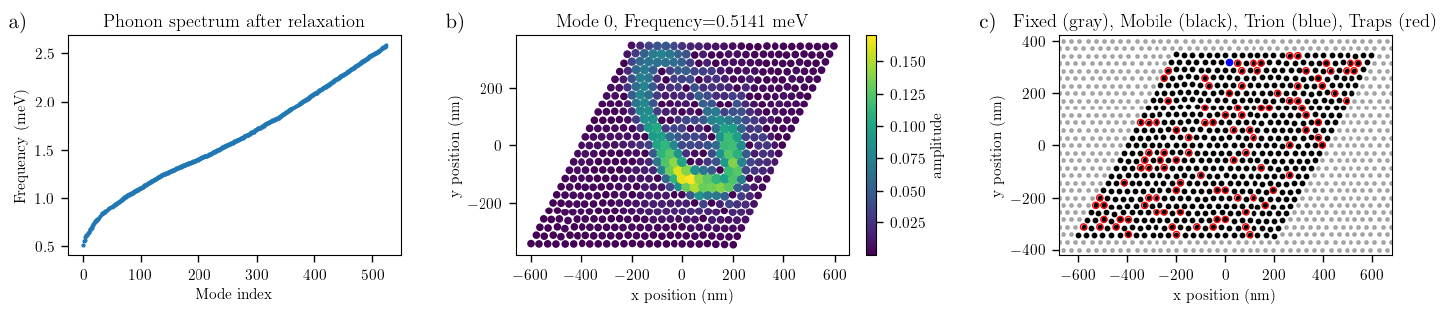}
\caption{
\textbf{Relaxed disordered Wigner crystal and its phonon spectrum.}
(a) Phonon eigenfrequencies as a function of mode index for a $25\times25$ triangular lattice with Yukawa interactions at $n = 10^{11} $ cm${}^2$. (b) The spatial distribution of the lowest energy eigenmode, illustrating the emergence of localized phonon modes away from the pinned regions. (c) Real-space configuration of the same system, showing fixed electrons (gray), mobile electrons (black), a single trion (blue) with larger mass, and pinned disorder sites (red) confined by harmonic traps.
}
\label{fig:S2_modes_and_configuration}
\end{figure}

\subsection{Disorder effects and phonon localization}

To explore how disorder affects collective excitations, we analyze the local density of states (LDOS),
\begin{align}
    A_j(\omega) = \sum_{n,\nu} |\langle n| j,\nu\rangle|^2\, \delta(\omega - \omega_n) \,,
\end{align}
at trion or electron sites $j$, averaged over multiple realizations of disorder and trion positions. In all LDOS spectra, we apply a numerical broadening of 0.05 meV to smooth the discrete phonon modes of the finite simulation cell.

\begin{figure}[t]
\centering
\includegraphics[width=0.8\textwidth]{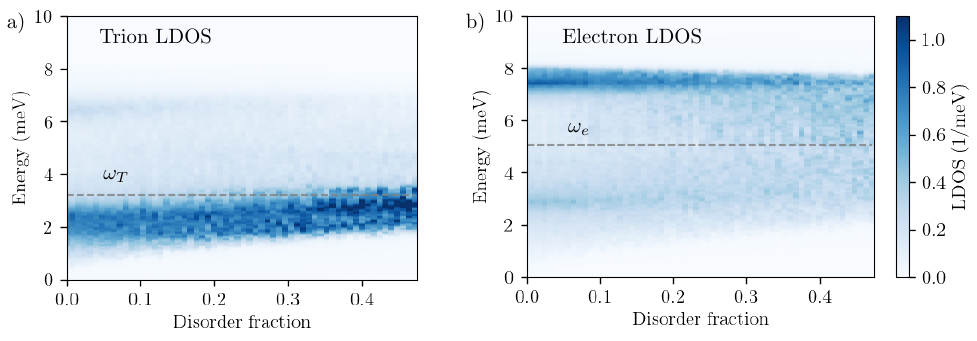}
\caption{
\textbf{Trion and electron LDOS as function of disorder fraction.}
(a) Local density of states (LDOS) at the trion site for increasing fraction of pinned disorder sites and fixed electron density. (b) LDOS at representative electron sites for the same disorder fractions. The dashed gray lines indicate the bare trion and electron frequency, $\omega_T$ and $\omega_e$, respectively. All spectra are averaged over 42 different trion or electron positions for a $25\times25$ lattice with lattice constant $a=20~\mathrm{nm}$.
}
\label{fig:S3_disorder_fraction}
\end{figure}

We first fix the electron density $n$ and vary the fraction $N_{\text{disorder}}/N$ of pinned disorder sites. Figure~\ref{fig:S3_disorder_fraction} shows the averaged LDOS at trion and electron sites for a constant electron density of $n= 2.89 \times 10^{11}$ cm${}^{-2}$, corresponding to $a=20~\mathrm{nm}$. In Fig.~\ref{fig:S3_disorder_fraction}a), the LDOS shows a dominant resonance around the expected bare trion frequency $\omega_T$ (dashed line), while in Fig.~\ref{fig:S3_disorder_fraction}b) the LDOS is peaked at the upper edge of the phonon continuum. The weak feature in Fig.~\ref{fig:S3_disorder_fraction}b) around 3 meV originates from an enhanced lattice density of states in the electron WC, and is also present in the absence of a trion. Interaction effects shift and broaden the trion resonance, reflecting the hybridization between local trion vibrations and extended WC phonons. Increasing disorder induces a small upward shift of the lowest phonon frequencies, consistent with pinning-induced localization of vibrational modes~\cite{chen:2025}.

For weak disorder ($N_{\text{disorder}}/N < 0.1$), the LDOS displays a broad resonance with a width that is primarily determined by the intrinsic phonon dispersion. Different disorder realizations produce nearly indistinguishable spectra, indicating that weak pinning perturbs the vibrational structure only minimally. In this regime, the spectral broadening is largely homogeneous, arising from the finite phonon bandwidth.

\begin{figure}[b]
\centering
\includegraphics[width=0.7\textwidth]{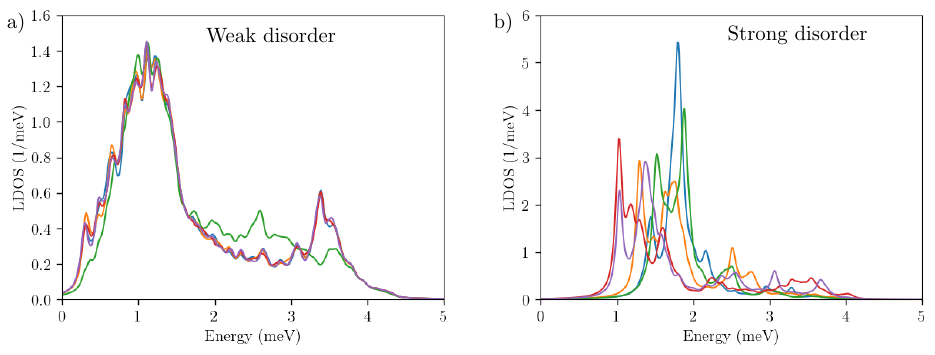}
\caption{
\textbf{Trion LDOS for weak and strong disorder.}
a) Trion LDOS for several different trion positions and weak disorder, $N_{\text{disorder}}/N=0.002$. b) Trion LDOS for several different trion positions and strong disorder, $N_{\text{disorder}}/N=0.5$. The broadening changes from homogeneous to inhomogeneous for stronger disorder, signifying a crossover from delocalized to localized phonon excitations.
}
\label{fig:S4_disorder_linecuts}
\end{figure}

At larger disorder fractions, the LDOS undergoes a clear qualitative change. The main resonance broadens asymmetrically and eventually splits into multiple sub-peaks, reflecting the emergence of local strain fields and spatial variations of the restoring potential. Representative line cuts for weak and strong disorder in Fig.~\ref{fig:S4_disorder_linecuts} illustrate this transition: weak disorder yields a broad peak dominated by the intrinsic phonon bandwidth, whereas strong disorder produces pronounced inhomogeneous broadening. This behavior is characteristic of localized vibrational excitations. Since each phonon mode experiences a slightly different local stiffness, its resonance frequency varies from site to site, giving rise to the multi-peaked, asymmetric spectra observed at high disorder.

Overall, disorder has a comparatively modest effect on the resonance frequency but a strong impact on the line shape, driving a crossover from homogeneous to inhomogeneous broadening as the number of pinned sites increases. To which degree the difference between homogeneous or inhomogeneous broadening of the phonons affects the polaron spectrum is an interesting open question.

\subsection{Density-dependence of the trion LDOS in presence of disorder}

We next study the LDOS at the trion site as a function of electron density $n$. In this analysis, we assume that disorder originates from a fixed set of impurities in the substrate, characterized by a disorder fraction of $N_{\text{disorder}}/N = 0.5$ at the reference electron density $n_0 = 10^{11}\,\mathrm{cm^{-2}}$. When varying the electron density $n$, we keep the total number of electrons $N$ fixed and adjust the simulation area accordingly. As a result, the fraction of pinned sites decreases with increasing $n$, according to $N_{\text{disorder}}/N = 0.5\, n_0/n$, so that local pinning effects are strongest at low electron densities.

\begin{figure}[b]
\centering
\includegraphics[width=0.78\textwidth]{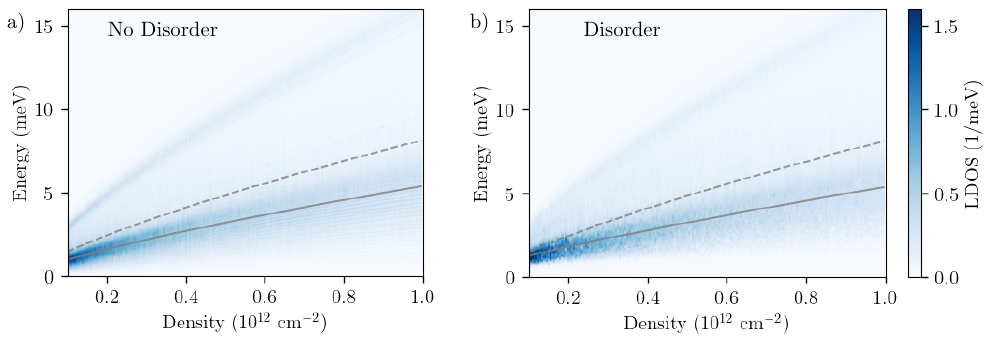}
\caption{
\textbf{Trion LDOS as function of electron density.}
Trion LDOS in case of (a) no disorder and (b) disorder, as described in the text. Dashed lines show the bare trion frequency $\omega_T(n)$ from Eq.~\eqref{eq:bare_trion_freq}. 
Solid lines show fits to the extracted peak position. The extracted parameters are a) $\omega_T(n) = (0.97 n^{0.75}+0.003)$ meV and b) $\omega_T(n) = (0.7 n^{0.84}+0.56)$ meV, where $n$ is in units of $10^{11}$ cm${}^{-2}$. All spectra are averaged over 42 different trion positions.
}
\label{fig:S5_trion_spectra}
\end{figure}

The resulting spectra, presented in Fig.~\ref{fig:S5_trion_spectra}, show that disorder leads to a slight upward shift of the lowest phonon frequencies and causes a transition from homogeneous to inhomogeneous broadening in the trion LDOS, consistent with a pinning-induced localization of vibrational modes. For each density, we extract the dominant resonance frequency $\omega_{\rm peak}(n)$ by fitting the LDOS. In the clean limit, the peak follows the expected WC phonon scaling,
\begin{equation}
    \omega_{\rm peak}(n) \propto n^{3/4} \,.
\end{equation}
In Fig.~\ref{fig:S5_trion_spectra}b), in presence of disorder, the peak of the trion LDOS only experiences a small shift compared to Fig.~\ref{fig:S5_trion_spectra}a). However, fitting the data yields a power-law exponent of 0.84 and a small offset at zero density. This indicates that even relatively small density-dependent changes in the data can produce substantial changes in the extracted fit parameters. In presence of strong disorder, the peak in the trion LDOS near $\omega_T$ does not persist down to zero density. At sufficiently low densities, all sites are pinned and no clear resonance in the trion LDOS is expected, since the pinning frequency is likely to be different per site, leading to strong inhomogeneous broadening.

Notably, the trion-mode frequency in the presence of disorder appears slightly more linear as a function of density, consistent with trends observed in experiment. Our findings demonstrate that even moderate disorder can substantially modify both the effective prefactor and the extracted exponent.

\edit{
\section{Comparison with recent publications}

As discussed in the main text, several related publications appeared around the same time as the present work. We provide here a brief comparison of the various theories presented in these works. These include a series of primarily experimental papers that report the first observation of Wigner polarons~\cite{zhang_2:2025,wang:2025,Liu2026}. In Ref.~\cite{zhang_2:2025}, the present authors contributed to the theoretical analysis of the experiment. The $n^{3/4}$ scaling of the Wigner polaron, as derived in Eq.~(3), was shown to be consistent with the experimental observations.  
 In addition, another theoretical work addressing the dressing of excitons by Wigner crystal phonons has recently appeared~\cite{Nyhegn2025}. 
 Here, we place the present work in the context of the theoretical developments in each of these works below.

\subsection{Ref.~\cite{wang:2025}: ``Spectroscopy of Wigner crystal polarons in an atomically thin semiconductor''}

Ref.~\cite{wang:2025} provides an alternative microscopic description of the Wigner polaron. They employ a Hartree-Fock inspired treatment where the Wigner crystal is represented by a periodic potential acting on the electrons, considering spin-polarized electrons. The electronic ground state is then taken to be a Fermi sea occupying the lowest Bloch band of this potential, while the elementary electronic excitations correspond to transitions into higher Bloch bands. Although the authors of Ref.~\cite{wang:2025} provide a different interpretation of these results, their calculation suggests a qualitatively similar picture to ours: the Wigner polaron emerges from the dressing of the attractive polaron by elementary electronic excitations, which in this case are particle--hole-like rather than phonon-like. In both cases the predominant contribution to the WP wave function would stem from an excited motional state of the trion.

The model from the present paper and from Ref.~\cite{wang:2025} differ in the predicted density dependence of the Wigner polaron energy. In Ref.~\cite{wang:2025}, the density dependence of the Wigner polaron energy is not extracted directly from the variational polaron calculations. This is firstly because the depth of the effective Wigner crystal potential the electrons experience is not calculated, but chosen by hand. Secondly, in performing the polaron calculation, a truncated basis including only three electronic Bloch bands was used. This effectively imposes a UV cutoff on the electronic momenta of order $1/a$, which is insufficient to capture the short-range correlations of the tightly bound trion at the relevant densities. 

Instead, the density dependence of the WP is fitted to the experiment with a two-level model which the authors motivate based on the previously described calculations. This phenomenological model 
consists of two states, which are argued to correspond to the ``dressed'' zero-momentum exciton and the ``dressed'' Umklapp exciton. Their energies are given by the bare exciton and the Umklapp exciton energies, shifted down by the trion binding energy. These states are then hybridized by an off-diagonal coupling which is taken to be the exciton-electron $T$-matrix evaluated at the Fermi energy. The lower and upper eigenstate of this model are then interpreted as the AP and WP, respectively. Within this picture, Wigner polaron energy is very non-linear at low densities due to the $1/\log(E_F)$ dependence of the $T$-matrix, while at high densities it becomes linear with a slope set by the Umklapp energy of the exciton. With a single fitting parameter, this model is in good agreement with the experimentally extracted Wigner polaron energies.

However, the exciton--electron $T$-matrix evaluated at the Fermi energy is associated with the scattering of an exciton with an electron at the Fermi momentum, and it is not clear why this should enter as the off-diagonal coupling in the effective two-level model. Furthermore, it is not obvious why the exciton-like states are the right basis to express this wave function. Indeed, even for the largest experimental density where the Wigner crystal appears of $7 \times 10^{11}$ cm$^{-2}$, the Fermi energy $E_F\approx 3.1$ meV is much smaller than the trion binding energy $E_T$ ($E_F \ll |E_T|)$. For non-interacting fermions, this regime therefore corresponds to the low-density regime of the Fermi polaron, where the AP has relatively little quasiparticle weight~\cite{schmidt:2012}, and therefore little free exciton content in the wave function. Our model finds that this is analogous in the Wigner crystal case, and that the WP also has low free-exciton content.
Instead, the ``tetron'' part of the wave function (consisting of a trion and a hole) is dominant, which is the regime we focused on in our calculations. 
In our model it is thus the phonon energy rather than the Umklapp energy which is the most important contribution to the splitting of the WP and the AP. For these reasons, the two-level model and its interpretation do not seem to be consistent with our theoretical description.

\subsection{Ref. \cite{Liu2026}: ``Exciton-polaron Umklapp scattering in Wigner crystals''}
Ref.~\cite{Liu2026} presents a primarily phenomenological model, in which the relevant quasiparticles are treated as discrete states that are coupled together by fitting parameters. In that work, the attractive polaron stems from a quasiparticle termed the tetron: a localized trion accompanied by a vacancy in the Wigner crystal. This picture of the attractive polaron is consistent with what we put forth in the present work. They argue that the Wigner polaron stems from the Umklapp of the tetron where its energy density dependence is set to match that of the exciton Umklapp. 

However, it is unclear from the microscopic arguments how the tetron Umklapp peak arises. In general, the Umklapp peak is caused due to the backfolding of the free particle dispersion, perturbatively interacting with a lattice potential. While for the repulsive polaron this is clearly a good picture, this cannot be applied to the tetron in the Wigner crystal since it is localized to a lattice site. While the tetron can form a delocalized polaron through hopping via an intermediate virtual exciton state, this will create only a single tetron band and therefore no Umklapp peak. Higher bands of the tetron in this scenario would only appear from excited motional states within each site forming their own bands. This is in line with the picture from our work. 

Aside from the microscopic interpretation, we highlight that the phenomenological model presented in Ref.~\cite{Liu2026} shows nice agreement with the experimental data. Also, their work considers other resonances, such as the Umklapp of the longitudinal exciton branch, which we do not consider in our work.

\subsection{Ref.~\cite{Nyhegn2025}: ``An exciton interacting with the phonons of an electronic Wigner crystal''}

Ref.~\cite{Nyhegn2025} considers a related, but distinct, problem of an exciton interacting with the phonon modes of a Wigner crystal in an adjacent layer. These calculations were carried out independently of the experimental observations and therefore address a different physical question rather than the Wigner polaron studied in the present work.

Since the exciton and Wigner crystal reside in different layers, the exciton has only weak attractive interactions with the electrons, giving rise to a polaron corresponding to a slightly redshifted exciton. The authors then describe the effect of the Wigner crystal phonons on the dressing of the exciton. Here, they go beyond the Hartree approximation in the present work, and describe the full phonon dispersion of the Wigner crystal. However, the exciton couples much more weakly to those phonons since the Wigner crystal resides in the other layer. Therefore they can capture the polaron properties via a diagrammatic approach within the self-consistent Born approximation. The Wigner polaron feature is absent in this model, because there is no tightly bound trion state.

}

\end{document}